# A user co-designed digital INtervention for Child LangUage DisordEr: The INCLUDE Project Protocol

Rafiah Ansari
National Institute of Health Research (NIHR) Clinical Doctoral Research Fellow, City University of London, UK
rafiah.ansari@city.ac.uk

**Abstract:**

Around ten percent of children may present with a disorder where language does not develop as expected. This often affects vocabulary skills, i.e., finding the words to express wants, needs and ideas, which can influence behaviours linked to wellbeing and daily functioning, such as concentration, independence, social interactions and managing emotions. Without specialist support, needs can increase in severity and continue to adulthood.

The type of support, known as interventions, showing strongest evidence for improving vocabulary with some signs of improved behaviour and wellbeing are ones that use word webs. These are diagrams consisting of lines that connect sound and meaning information about a word to strengthen the child's word knowledge and use. The diagrams resemble what is commonly known as mind-maps and are widely used by Speech and Language Therapists in partnership with school educators to help children with language difficulties. In addition, interventions delivered through mobile-devices has led in some cases to increased vocabulary gains with positive influence on wellbeing and academic attainment.

With advances in technology and availability of user-friendly mobile devices to capture, combine and replay multimedia, new opportunities for designing bespoke vocabulary instruction have emerged that are without timing and location constraints. This brings the potential to engage and motivate users and harbour independence through functional strategies that support each child's unique language needs. To achieve this, children with language disorder, their parents/carers, support professionals and software development team members must work jointly to create an intervention that is fit for purpose. This is the first research planned to explore the collaborative development and acceptability of a digitally enhanced vocabulary intervention for child language disorder.

**CCS Concepts:**

Human-Centred Computing → Empirical studies in Human Computer Interaction, Accessibility, and Collaborative and Social Computing

**Additional Key Words and Phrases:**

Therapeutic technology, Speech and Language Therapy, Child Language Disorder, Vocabulary intervention, Word learning

**ACM Reference Format:**

Rafiah Ansari. 2023. A User Co-designed Digital Intervention for Child Language Disorder: The INCLUDE Project Protocol. Paper has been selected following peer review for presenting at the "CHI 2023 Workshop on Child-centred AI Design: Definition, Operation and Considerations, April 23, 2023, Hamburg, Germany."

The study reported in this paper has been granted ethics approval by the City, University of London Research Ethics Committee.

---

## 1 INTRODUCTION

Language disorder, a neurodevelopmental condition characterised by persistent and severe challenges in daily communication that requires specialist intervention (Bishop, et al., 2017), is estimated to affect one in ten children (Norbury et al., 2016). Whilst the disorder takes many forms, difficulty with word-finding, where children struggle to find the right word to express themselves, is a common feature (Dockrell et al., 1998). It is unsurprising that identification of needs is most common in the primary school years, i.e. age 5-11 years (Lindsay and Strand, 2016), given that during this period there is an expected estimated vocabulary growth from 3, 000 to 8, 000 words (Anglin et al., 1993; Biemiller and Slonim, 2001). Vocabulary difficulties rarely occur in isolation and have been found to predict both wider language competencies and literacy skills as well as forecasting weak educational attainment from the primary school years through to adolescence (Friend et al., 2019; Reilly et al., 2010; Duff et al., 2015; Oslund et al., 2018; Bleses et al., 2016). Around a third to half of children with language disorder characterised by weak vocabulary also present with behaviour needs that impact on wellbeing including inattention, social difficulties, anxiety and conduct issues (Maggio et al., 2014; Tomblin et al., 2000; Willinger et al., 2003).

Intervention for children with language disorder align with the Medical Research Council's (MRC) definition of 'complex intervention' (Skivington et al, 2021) as it involves multiple professionals, settings and timeframes. Therapy is predominantly led by Speech and Language Therapists (SLTs) who deliver evidence-based therapy in conjunction with parents, educators and psychologists in a range of settings such as the home, school or clinic and with a range of language goals including those relating to the child's interest, curriculum focus and family routine (Law, Dennis and Charlton, 2017).

Intervention studies showing greatest potential for improving vocabulary with indications of behaviour gains are those employing word webs i.e., diagrams resembling mind-maps for organizing and forming connection connections between sound and meaning cues about a word (Best et al., 2021). Word web intervention is widely used by SLTs to help children with language disorder (Best et al., 2018) and the benefits of word web therapy have been demonstrated through case studies, randomised crossover studies and randomised control trails (Best et al., 2015, 2018, 2021; St John and Vance, 2014).

Existing theory and research around language acquisition indicates that the word web method of strengthening connections between word sound and word meaning cues is more robust when dynamic multisensory, multitemporal cues are used rather than static unimodal input (Yu and Ballard, 2007; Monaghan, 2017). Hollich et al (2000) propose a coalition language acquisition model where multiple sources, such as perceptual salience, prosodic cue, eye gaze, social context, syntactic cues, and temporal contiguity, combine. Adopting user-friendly mobile devices to capture, combine and replay multimedia content (i.e., text, image, audio and video) has been found to support the cross-situational, multimodal cue integration required for optimum word learning in children with language disorder. Lowman & Dressler (2016) demonstrated that repeatedly viewing pre-recorded audio and video media around key vocabulary via an iPod in addition to reading a related topic book significantly increased understanding and use of target words in children aged 10-11 years with developmental language disorders compared to books alone. The author's own research found that late-talking children aged 7-11 repeated new word-forms more accurately if they saw them spoken onscreen compared to just hearing them (Badat, 2014). Interestingly, typically developing controls did not gain accuracy from visual-cues yet still outperformed the late-talkers indicating that multi-sensory input may be of particular benefit when learning novel word-forms for school-aged children with weak language-learning skills.

The potential for the multimodal, multitemporal word learning afforded by digital to support children with language disorder is reinforced by studies comparing the use of digital devices in relation to children without a language disorder which report comparable length of time spent on devices and comparable or greater levels of socially-driven motivation (Blom et al. 2017, Durkin et al. 2010). Additional factors such as accessibility, portability, increased independence, motivation and engagement are also cited in the literature as reasons for embracing mobile digital technology to support children with learning difficulties (Draper Rodríguez et al. 2014, Ferdig et al. 2016).

A benefit of a digitally-enhanced word web method that enables children to record and review multisensory cues about words from their own environment in their own time, is the potential to engage and motivate users and harbour independence whilst encouraging functional strategies to support each child's unique language needs. For this to be achieved, children with language disorder, their parents/carers, SLTs and Educators need to participate in research to jointly design and develop an intervention tool that is fit for purpose. As yet, there has been no research exploring the collaborative design and development of a digital word web intervention to support vocabulary leaning for child language disorder.

## 2 STUDY AIMS

- This study will involve co-design sessions with children who have language disorder, their parents/carers, support professionals and a software designer team to co-create a digitally-enhanced vocabulary intervention prototype.
- A focus group will then be held with the parent/carers and support professionals to determine the acceptability of the intervention prototype and further refinement will be carried out based on the feedback.

## 3 METHOD

### 3.1 Design

This study will adopt a participatory research approach using co-design sessions and focus groups. The term 'participatory research' refers to both a series of techniques and an ideological perspective which enables subjects of research to become involved as partners in the process of the enquiry (Slattery et al, 2020).

This is achieved by encompassing research designs, methods, and frameworks that use systematic inquiry in direct collaboration with those affected by the issue being studied (Vaughn and Jacquez, 2020).

#### *3.1.1 Co-design*

The study will consist of a series of co-design workshops spread over a school term (12 weeks) involving two groups of children with language disorder from two primary schools, their parents/carer, school staff that are familiar with the children, each school's SLT, the primary researcher and a software design team. The purpose of the workshops will be to map the user experience (Szabo, 2017), i.e., to gain a complete picture of how the children with language disorder may interact with a digitised version of a word web.

This learning will help to understand and prioritise the user needs whilst identifying the digital intervention features that require refinement. This will allow the creation of a user co-designed intervention prototype that is relevant, appropriate and engaging when supporting children with language disorder. The specific stages for planning, running and reviewing the co-design workshops are detailed in Table 1.

#### *3.1.2 Focus group*

Once refinement saturation in the co-design sessions is reached and children are happy with the intervention prototype, the co-designed intervention prototype will be shared with the multidisciplinary stakeholder group (parents/carers, education staff, SLTs).

This will involve running a focus group on the acceptability of the prototype as a language intervention tool measured using the empirically validated Theoretical Framework of Acceptability (TFA) (Sekhon, Cartwright & Francis, 2022). Focus group findings on intervention acceptability will guide further refinements of the digital prototype.

### 3.2 Accessibility

Given that the population of focus are children presenting with communication needs, this study will incorporate best practice guidance for inclusive and accessible participatory research. In 2019, the National Institute of Health Research (NIHR), which is the largest funder of UK health and care research, introduced the UK Standards for Public Involvement, a six-part framework for what good participatory healthcare research should look like (Figure 1).

In 2020, the Royal College of Speech and Language Therapists (RCSLT), the professional body for SLTs in the UK, adopted and advocated use of the UK Standards for Public Involvement [NIHR, 2019) to provide guidance, tools and techniques for participatory research involving children who have language disorder, their carers and support professionals (Chadd et al., 2020). The recommendations and standards will be incorporated into the design of the study as detailed in Table 1.

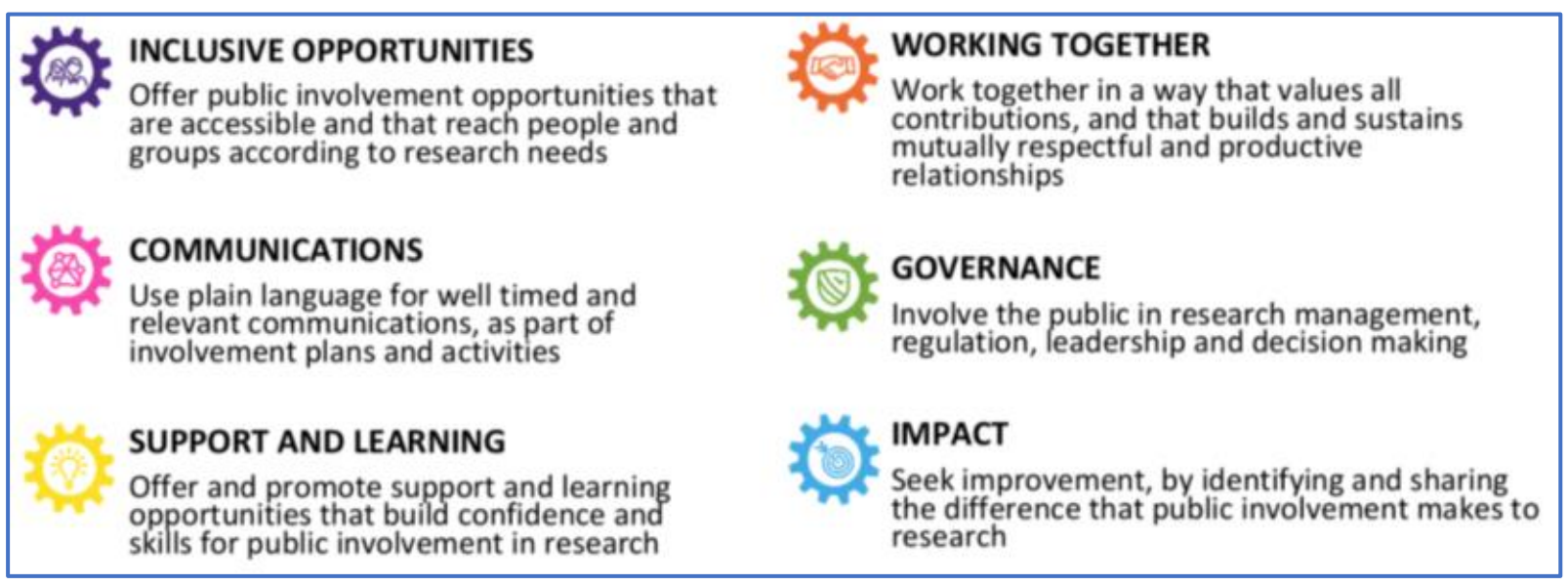

**Figure 1**: UK Standards for Public Involvement (NIHR, 2020 - page 4 of slide set)
uk-standards-for-public-involvement_full-slide-set_nov19.pptx - google drive

**Table 1: Step by step plan for co-design workshops**

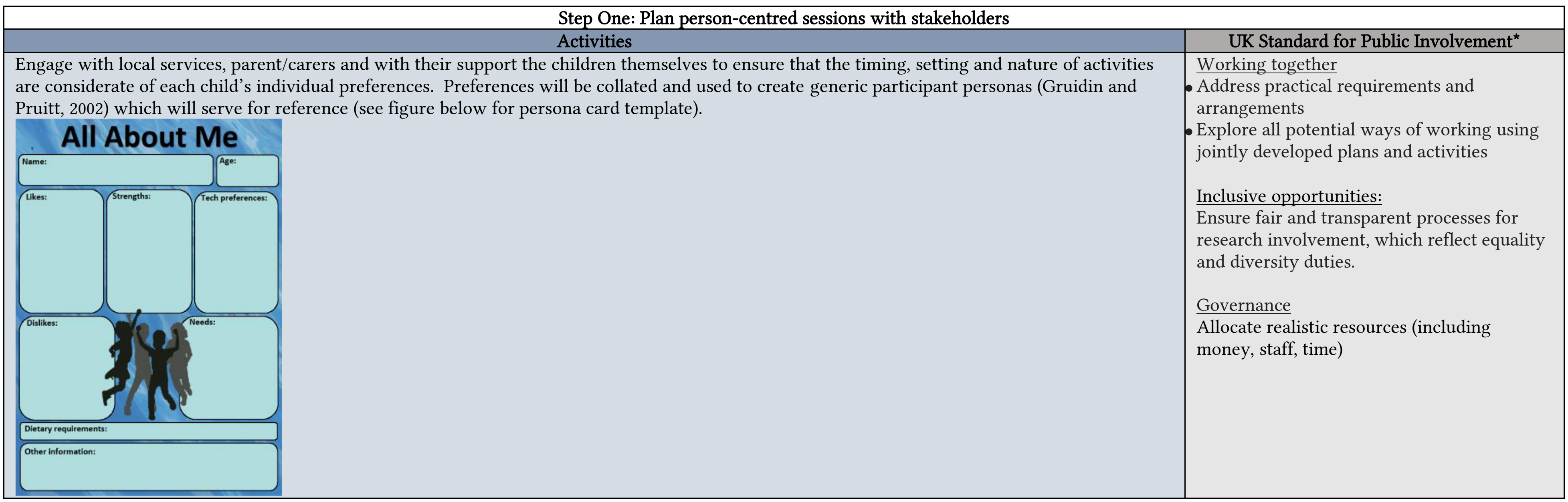

| Step One: Plan person-centred sessions with stakeholders | |
|---|---|
| **Activities** | **UK Standard for Public Involvement*** |
| Engage with local services, parent/carers and with their support the children themselves to ensure that the timing, setting and nature of activities are considerate of each child's individual preferences. Preferences will be collated and used to create generic participant personas (Gruidin and Pruitt, 2002) which will serve for reference (see figure below for persona card template). | <u>Working together</u><br>• Address practical requirements and arrangements<br>• Explore all potential ways of working using jointly developed plans and activities<br><br><u>Inclusive opportunities:</u><br>Ensure fair and transparent processes for research involvement, which reflect equality and diversity duties.<br><br><u>Governance</u><br>Allocate realistic resources (including money, staff, time) |

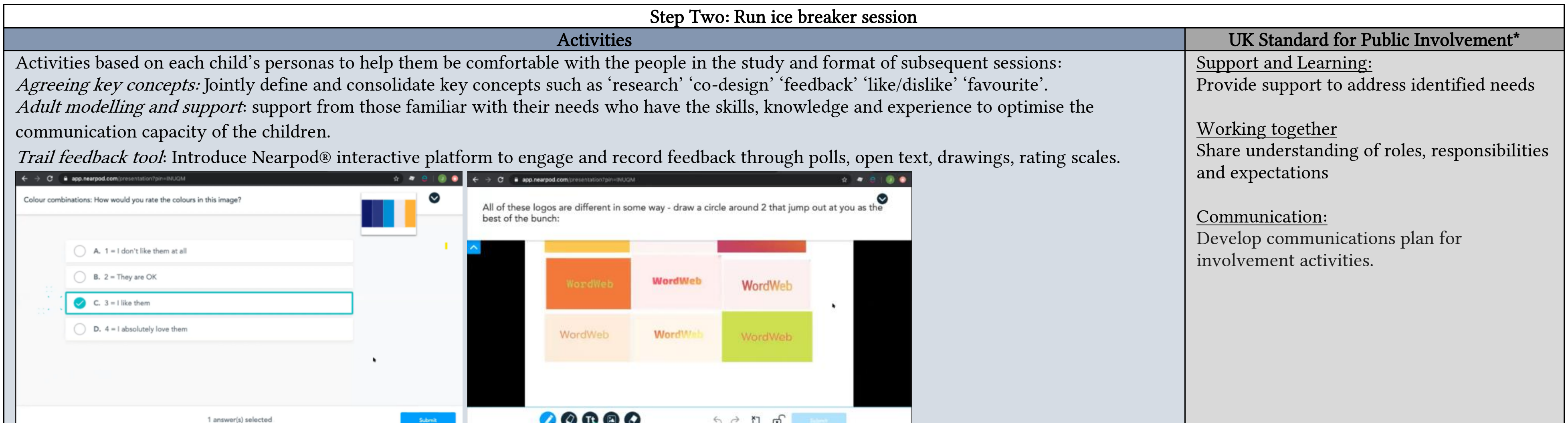

| Step Two: Run ice breaker session | |
|---|---|
| **Activities** | **UK Standard for Public Involvement*** |
| Activities based on each child's personas to help them be comfortable with the people in the study and format of subsequent sessions:<br>*Agreeing key concepts:* Jointly define and consolidate key concepts such as 'research' 'co-design' 'feedback' 'like/dislike' 'favourite'.<br>*Adult modelling and support*: support from those familiar with their needs who have the skills, knowledge and experience to optimise the communication capacity of the children.<br>*Trail feedback tool*: Introduce Nearpod® interactive platform to engage and record feedback through polls, open text, drawings, rating scales. | <u>Support and Learning:</u><br>Provide support to address identified needs<br><br><u>Working together</u><br>Share understanding of roles, responsibilities and expectations<br><br><u>Communication:</u><br>Develop communications plan for involvement activities. |

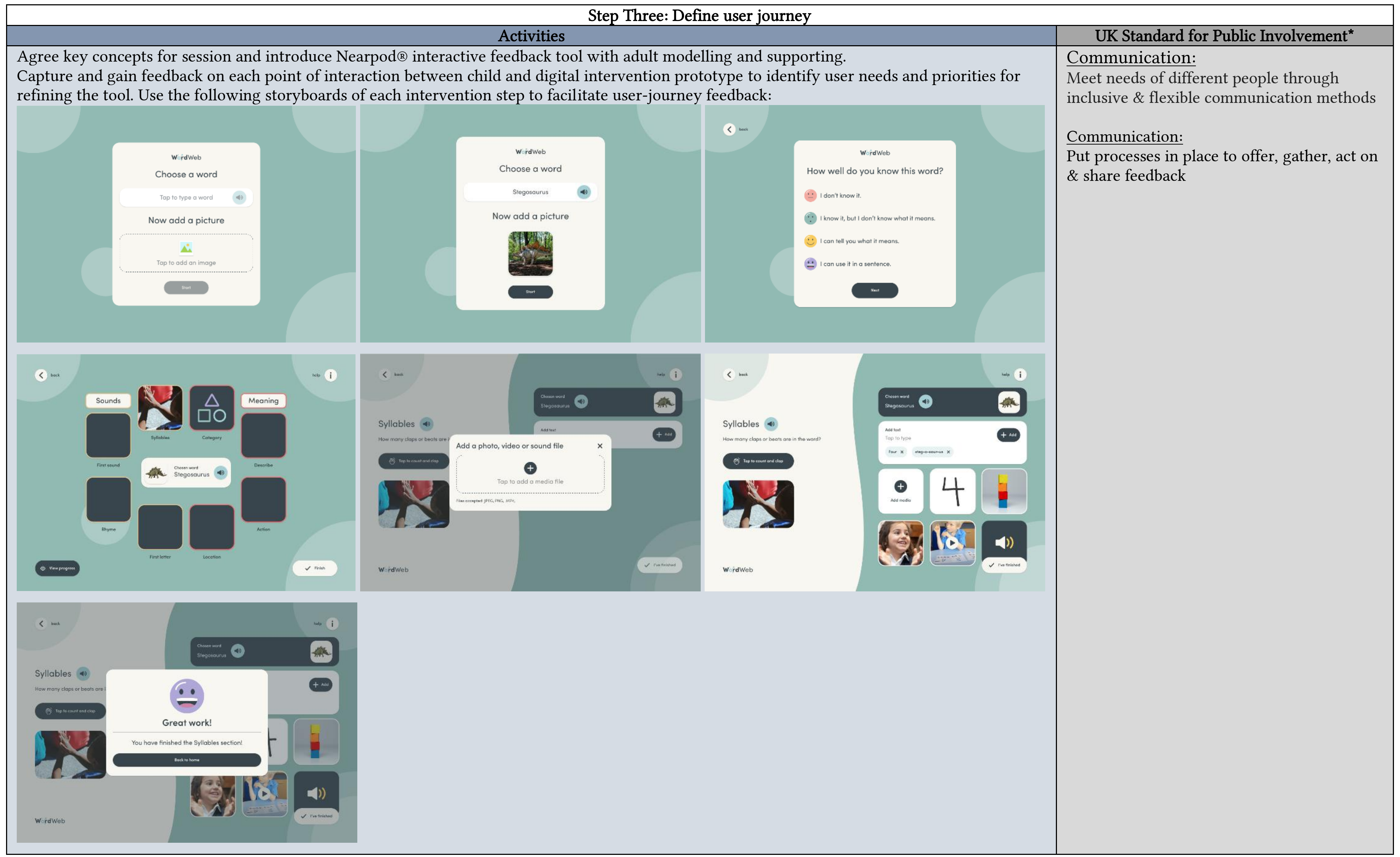

**Step Three: Define user journey**

| Activities | UK Standard for Public Involvement* |
| --- | --- |
| Agree key concepts for session and introduce Nearpod® interactive feedback tool with adult modelling and supporting.<br>Capture and gain feedback on each point of interaction between child and digital intervention prototype to identify user needs and priorities for refining the tool. Use the following storyboards of each intervention step to facilitate user-journey feedback: | <u>Communication:</u><br>Meet needs of different people through inclusive & flexible communication methods<br><br><u>Communication:</u><br>Put processes in place to offer, gather, act on & share feedback |

| Step Four: Address user needs (this may take multiple sessions) | |
|---|---|
| **Activities** | **UK Standard for Public Involvement*** |
| Agree key concepts for session and introduce Nearpod® interactive feedback tool with adult modelling and supporting.<br>Focus on specific scenarios of a user's interaction based on change priorities identified during the user journey (Step Three). Provide alternate designs to identify best solution. | Impact<br>Act on changes, benefits and learning resulting from co-involvement |

| Step Five: Ensure acknowledgement of contribution | |
|---|---|
| **Activities** | **UK Standard for Public Involvement*** |
| Certificate of achievement and gratitude vouchers provided regardless of duration of participation. This will be arranged in conjunction with each child's school and parent/carer. In addition, information on participation will be shared with permission on the school and SLT services websites and newsletters in accessible formats. | Communication:<br>Share public involvement learning and achievements<br><br>Working together<br>Recognise individuals' influence, ideas and contributions |

*UK Standard for Public Involvement https://sites.google.com/nihr.ac.uk/pi-standards/standards?authuser=0

### 3.2 PARTICIPANTS

1. Children of UK upper primary school (age 7-10 years) with diagnoses of language disorder or the equivalent according to criteria in Bishop et al. (2017).
2. Stakeholders: Given the multi-disciplinary, multi-situational model for child language disorder interventions (Dennis, Law and Charlton, 2017), the co-design process will be guided by the following multidisciplinary stakeholder group representing those who would be involved in implementing the novel intervention in practice:
    - Parents/carers of children with language disorder
    - Education staff supporting children with language disorder (this may differ slightly for each school but usually consists of Special Educational Needs Coordinator/Inclusion Managers, Teachers and Teaching Assistants)
    - SLTs supporting children with language disorder in primary school settings.

### 3.3 Sampling

Given the scope of the study and resources available (time, funding, research staff), this study will limit to two schools and aim to recruit four to six children with DLD, one SLT, and one educator (School SENCO, class teacher or teaching assistant) within each school. All sessions will be simultaneously replicated in each school, if there is marked discrepancy in co-design session feedback and findings with regards to prototype refinements then consensus will be sought from the stakeholder group.

### 3.4 Data Analysis

#### *3.4.1 Co-design*

Feedback from each co-design session as captured in the Nearpod® tool, video recordings and transcripts of recordings will be analysed identify and agree priorities for refining the intervention.

#### *3.4.2 Focus Group*

Once refinement saturation in the co-design sessions is reached, a focus group on the acceptability of the prototype as a language intervention tool will be held with the multidisciplinary stakeholder group (parents/carers, education staff, SLTs). Acceptability will be measured using the empirically validated theoretical framework of acceptability (TFA) (Sekhon, Cartwright & Francis, 2022). Acceptability in this context is defined as a multi-faceted paradigm that reflects the extent to which people delivering or receiving a healthcare intervention consider it to be appropriate, based on anticipated or experienced cognitive and emotional responses to the intervention (Sekhon et al., 2017). The TFA gages seven component constructs of acceptability: affective attitude (feelings about taking part in an intervention), burden (effort required to participate in the intervention), perceived effectiveness (likelihood for the intervention to achieve its purpose), ethicality (fit with value system), intervention coherence (understanding the intervention and how it works), opportunity costs (what must be given up to engage in an intervention), and self-efficacy (confidence that the behaviours required to participate in the intervention can be performed).

The TFA framework will guide the focus-group interview schedule and feedback from the focus group will be analysed according to the TFA constructs to identify priorities for further intervention refinement to maximise intervention acceptability. The software design team will update the prototype according to the stakeholder acceptability feedback.

### ACKNOWLEDGMENTS

The author, Rafiah Patel (Clinical Doctoral Research Fellow, ICA-CDRF-2018-04-ST2-029), is funded by the National Institute for Health Research (NIHR)/Health Education England (HEE) for this research project. The views expressed in this publication are those of the author and not necessarily those of the NIHR, NHS or the UK Department of Health and Social Care.